\documentclass[a4paper,12pt]{article}
\usepackage{times}
\usepackage{latexsym} % \Box etc ...
\usepackage{amsmath}
\usepackage{amssymb}
\usepackage{graphicx}
\usepackage{wrapfig}
\usepackage{sectsty}
\usepackage{indentfirst}

\allsectionsfont{\centering}
\makeatletter
\def\@seccntformat#1{\csname the#1\endcsname.\quad}
\makeatother

\newcommand{\be}{\begin{equation}}
\newcommand{\ee}{\end{equation}}
\newcommand{\ba}{\begin{eqnarray}}
\newcommand{\ea}{\end{eqnarray}}
\newcommand{\pa}{\partial}
\let\f\frac

\begin{document}

\title{\large\textbf{On gravitational flow in the Relativistic Theory of Gravitation}}
\date{}
\author{\normalsize S.\,S.~Gershtein, A.\,A.~Logunov, M.\,A.~Mestvirishvili}
\date{}
\maketitle

\begin{abstract}
A definition of the gravitational flow  and a short description of
the recipe of its calculation  are presented.
\end{abstract}

 In work [1] within the limits of the Relativistic Theory of Gravitation (RTG) the author has come to conclusion,
 that \textit{``the flow of gravitational radiation coming from any spatially limited source is  positive definite''}.
In this article we give a brief account of  this approach.

The system of RTG equations is as follows
\be
R^{\mu\nu}-\f{\,1\,}{2}g^{\mu\nu}R+\f{m^2}{2}
\Bigl[g^{\mu\nu}+\Bigl(g^{\mu\alpha}g^{\nu\beta}
-\f{\,1\,}{2}g^{\mu\nu}g^{\alpha\beta}\Bigr)\gamma_{\alpha\beta}\Bigr]
=8\pi T^{\mu\nu}\,,
\label{eq1}
\ee
\be
D_\mu\tilde{g}^{\mu\nu}=0\,.
\label{eq2}
\ee
These equations are generally covariant concerning any coordinate transformations and form-invariant
concerning the transformations leaving Minkowski metric
$\gamma_{\mu\nu}(x)$ unchanged. The effective metric of Riemannian space
$g^{\mu\nu}$ is related to the gravitational field $\phi^{\mu\nu}$ by equation
\be
\tilde{g}^{\mu\nu}=\tilde{\gamma}^{\,\mu\nu}+\tilde{\phi}^{\mu\nu},
\label{eq3}
\ee
where $\tilde{g}^{\mu\nu}=\sqrt{-g}\,g^{\mu\nu}$, $\tilde{\gamma}^{\,\mu\nu}=
\sqrt{-\gamma}\,\gamma^{\mu\nu}$, $\tilde{\phi}^{\mu\nu}=\sqrt{-\gamma}\,\phi^{\mu\nu}$;
$\gamma_{\mu\nu}$ is Minkowski space metric, $g=\det g_{\mu\nu}$,
$\gamma=\det \gamma_{\mu\nu}$.

For the further consideration it is necessary for us to write down Eq.~(\ref{eq1}) in the following form
\be
D_\alpha D_\beta (\tilde{\phi}^{\alpha\beta}\tilde{\phi}^{\varepsilon\lambda}
-\tilde{\phi}^{\varepsilon\beta}\tilde{\phi}^{\lambda\alpha})
=-\tilde{\gamma}^{\alpha\beta}D_\alpha D_\beta\tilde{\phi}^{\varepsilon\lambda}
-m^2\sqrt{-\gamma}\,\tilde{\phi}^{\varepsilon\lambda}
-16\pi g(T^{\varepsilon\lambda} +t_g^{\varepsilon\lambda}),
\label{eq4}
\ee
where $T^{\varepsilon\lambda}$ is the energy-momentum tensor of substance,
$t_g^{\varepsilon\lambda}$ is the tensor of the  gravitational field.
\newpage
\ba
&&\!\!\!\!\!\!\!\!\!\!-16\pi g t^{\varepsilon\lambda}_{g}=\frac{\,1\,}{2}
\Bigl(\tilde g^{\varepsilon\alpha} \tilde g^{\lambda\beta}-\frac{\,1\,}{2}
\tilde g^{\varepsilon\lambda} \tilde g^{\alpha\beta}\Bigr)
\Bigl(\tilde g_{\nu\sigma}\tilde g_{\tau\mu}-\frac{\,1\,}{2}
\tilde g_{\tau\sigma}\tilde g_{\nu\mu}\Bigr)\times
\nonumber  \\
&&\!\!\!\!\!\!\!\!\!\!\times D_\alpha \tilde\phi^{\tau\sigma}D_\beta
\tilde\phi^{\mu\nu}
+\tilde g^{\alpha\beta}\tilde g_{\tau\sigma} D_\alpha
\tilde\phi^{\varepsilon\tau} D_\beta \tilde\phi^{\lambda\sigma}
\!+\!\frac{\,1\,}{2}\tilde g^{\varepsilon\lambda}
\tilde g_{\tau\sigma} D_\alpha \tilde\phi^{\sigma\beta}
D_\beta \tilde\phi^{\alpha\tau}
-\!\!
\nonumber  \\
&&\!\!\!\!\!\!\!\!\!\!-\tilde g^{\varepsilon\beta} \tilde g_{\tau\sigma}D_\alpha
\tilde\phi^{\lambda\sigma} D_\beta\tilde\phi^{\alpha\tau}-\tilde g^{\lambda\alpha} \tilde g_{\tau\sigma}D_\alpha
\tilde\phi^{\beta\sigma}D_\beta
\tilde\phi^{\varepsilon\tau}-
\nonumber  \\
&&\!\!\!\!\!\!\!\!\!\!
-m^2
\Bigl(
\sqrt{-g}\tilde g^{\varepsilon\lambda}
-\sqrt{-\gamma} \tilde\phi^{\varepsilon\lambda}
+\tilde g ^{\varepsilon\alpha} \tilde g^{\lambda\beta}
\gamma_{\alpha\beta}-\frac{\,1\,}{2} \tilde g^{\varepsilon\lambda}
\tilde g^{\alpha\beta}\gamma_{\alpha\beta}
\bigl. \Bigr)\,,
\label{eq5}
\ea
here $D_\alpha$ is the covariant derivative in the  Minkowski space-time.

As Eq.~(\ref{eq4}) is considered as the field equation in Minkowski space, the order of derivatives is insignificant. Rise and lowering of indicies is carried out by   metric tensor $\gamma_{\mu\nu}$. So, for example, if the contravariant momentum of a graviton is
\[
p^\mu = m\f{dx^\mu}{ds}\,,
\]
then the covariant momentum is given by formula
\[
p_\nu = \gamma_{\nu\mu} p^\mu.
\]
From here it follows that
\[
g_{\mu\nu} p^\mu p^\nu = m^2,
\]
whereas
\[
g^{\mu\nu} p_\mu p_\nu \neq m^2.
\]
It means that
\be
ds^2=g_{\mu\nu}dx^\mu dx^\nu,
\label{eq6}
\ee
whereas
\be
ds^2 \neq g^{\mu\nu}dx_\mu dx_\nu,\;\;\mbox{here}\;\; dx_\mu =\gamma_{\mu\nu}dx^\nu, \label{eq7}
\ee
at the same time
\[
d\sigma^2 = \gamma_{\mu\nu}dx^\mu dx^\nu = \gamma^{\mu\nu}dx_\mu dx_\nu\,.
\]
\textit{It follows from the above that the contravariant tensor $g^{\mu\nu}$ in our description does not have metric properties}, though it is defined by the following equation
\[
g^{\mu\alpha} g_{\alpha\nu} = \delta_\nu^\mu.
\]
\textit{At our description metric properties are carried out only by a covariant metric tensor $g_{\mu\nu}$}.

In the following we provide all consideration in an inertial reference system at Galilean coordinates. The RTG system of  equations takes a form
\be
\pa_\alpha \pa_\beta
(\phi^{\alpha\beta}\phi^{\varepsilon\lambda}
-\phi^{\varepsilon\beta}\phi^{\lambda\alpha})
=-\gamma_{\alpha\beta}\pa^\alpha \pa^\beta\phi^{\varepsilon\lambda}
-m^2\phi^{\varepsilon\lambda}-16\pi g(T^{\varepsilon\lambda} +t_g^{\varepsilon\lambda})\,,
\label{eq8}
\ee
\be
\pa_\mu \phi^{\mu\lambda}=0\,,
\label{eq9}
\ee
here $\pa^\lambda=\gamma^{\lambda\nu}\pa_\nu$.

 A pecularity of the geometrized theory of gravitation, both  GRT, and RTG, is that the density of energy-momentum tensor of the gravitational field $t^{\varepsilon\lambda}$, defined according to Hilbert as the variation of the Lagrangian density of the gravitational field over  metric tensor
 $g_{\mu\nu}$, unlike in different theories, is precisely equal to zero outside the source as it is just the gravitational field equation.
 But from here it does not follow, that there is no gravitational radiation in the theory. As  $t^{\varepsilon\lambda}$ is the only general tensor characteristic of the second rank field outside the  source, it is natural to pick out the part of $t^{\varepsilon\lambda}$  responsible for a gravitational flow in a wave zone.

The density of energy-momentum tensor $t^{\varepsilon\lambda}$ has the following form
\be
16\pi\sqrt{-g}\,t^{\varepsilon\lambda}
=-\gamma_{\alpha\beta}\pa^\alpha \pa^\beta \phi^{\varepsilon\lambda}
-m^2\phi^{\varepsilon\lambda}
-16\pi g t_g^{\varepsilon\lambda}
-\pa_\alpha \pa_\beta (\phi^{\alpha\beta}\phi^{\varepsilon\lambda}
-\phi^{\varepsilon\beta}\phi^{\lambda\alpha})
\,.
\label{eq10}
\ee
Energy-momentum and  angular momentum conservation laws determine energy-momentum tensor up to some Krutkov tensor having identically zero divergence [3].  The fourth term in r.h.s. of Eq.~(\ref{eq10}) just represents a special  version of  the Krutkov tensor density. In a geometrized theory of gravitation the density of  Krutkov tensor arises from  Eqs.~(\ref{eq1}),(\ref{eq2}).  The tensor density in r.h.s. of  Eq.~(\ref{eq10}) consists of two parts: the first three terms are the first part  having zero divergence due to equations
(\ref{eq8}),(\ref{eq9}),  the second part includes density of the Krutkov tensor which is zero identically.
Just for this reason the   Krutkov tensor density itself does not reflect movement of  matter, but enters into the gravitational equation  in the certain special form. Thus, only the first part of the tensor density  determines the gravitational flow in a wave zone. Just  this part will be exploited by us, and the flow determined by it will be designated as  $J^i$.

 The well-known A.~Einstein's expression  for a gravitational radiation flow,  in the case when a graviton mass is equal to zero, could be obtained from   $J^i$  estimated in a wave zone on a solution of  equation
\be
\gamma_{\mu\nu}\pa^\mu \pa^\nu \phi^{\varepsilon\lambda}=0\,.
\label{eq11}
\ee
The solution is usually calculated  by means of the standard perturbation theory.
Following this procedure for  finding the gravitational radiation flow  in RTG it would be necessary to estimate  $J^i$ on a solution of the wave equation
\be
\gamma_{\mu\nu}\pa^\mu \pa^\nu \phi^{\varepsilon\lambda}+m^2\phi^{\varepsilon\lambda} =0\,,
\label{eq12}
\ee
which will be obtained according to the usual perturbation theory. But whether it is correct  to use this perturbation theory  in this case?

 According to wave equation  (\ref{eq12}) the gravitational wave propagates just in the Minkowski space because of presence of  the Minkowski space tensor $\gamma_{\mu\nu}$  in front of the second derivatives. But in fact it is not so even for a linear approximation because due to the effect of  gravitational field the metric tensor of  effective Riemannian space is given as follows
\be
g_{\mu\nu}
=\gamma_{\mu\nu}-\phi_{\mu\nu}+\f{\,1\,}{2}\phi\gamma_{\mu\nu},
\quad \phi =\phi_{\mu\nu}\gamma^{\,\mu\nu},
\label{eq13}
\ee
and therefore the propagation  of a gravitational wave occurs in the Riemannian space with scalar curvature $R$ which is
\[
R =\f{\,1\,}{2}m^2\phi.
\]
It means, that propagation of the gravitational wave in a wave zone occurs not according to  the wave equation (\ref{eq12})  of Minkowski space, but according to the wave equation of  Riemannian space with metric $g_{\mu\nu}$:
\be
g_{\mu\nu}\pa^\mu \pa^\nu \phi^{\varepsilon\lambda}+m^2\phi^{\varepsilon\lambda}=0\,.
\label{eq14}
\ee
In this equation changes in the metric due to the effect of a gravitational field are taken into account.
The effect of correction terms linear in the field and standing in front of the second derivatives cannot
be calculated in the standard perturbation theory.
But just by taking into account these terms we are able to avoid  the 
negative energy radiation
which usually occurs in the linear theory of a tensor field with nonzero graviton rest mass.
We will see this below.

 It has been shown in  [2] that already the second approximation of the perturbation theory  in  GRT
 can be \textit{``arbitrarily rising as opposed to the assumption made in  the approximation scheme.''
 \ldots `` Thus  the second approximation of $g_{ik}$ contains, except for periodic terms, as well terms
 quadratically rising with $x$. In higher approximations there are terms  even of higher orders of rising''}.
 On this basis C.~M\o ller has drawn a conclusion, that \textit{``\,`weak field approximation\!'
 is unsuitable for a study of such distributed solutions of the field equations as gravitational waves''}.
 For this reason it is necessary to use  the standard
 perturbation theory with caution in calculating the gravitational radiation flow, especially in  case
 when we are  to consider effect of a (weak also)
 gravitational field  on a variation of the space-time metric.

On the basis of all above-stated it follows that \textit{it is
necessary to calculate gravitational flow  $J^i$  in a wave zone
not on a  solution of  Eq.~(\ref{eq12}) which follows from the
perturbation theory, but on a solution of  wave equation
(\ref{eq14}) in which effect of a gravitational field on the
propagation of a wave is taken into account}. In case of zero
graviton rest mass the use of  Eq.~(\ref{eq14}) instead of
Eq.~(\ref{eq12}) gives the same result as Eq.~(\ref{eq12}).

Thus, a density of the flow calculated in a wave zone on a
solution of  Eq.~(\ref{eq14}) is equal to \be 16\pi J^i
=-\phi_{\alpha\beta}\pa^\alpha \pa^\beta\phi^{0i}
+\f{\,1\,}{2}\phi \gamma_{\alpha\beta}\pa^\alpha
\pa^\beta\phi^{0i} -16\pi g t_g^{0i}\,. \label{eq15} \ee From here
a total flow of gravitational radiation is as follows \be
J=\!-\!\!\!\oint\limits_{s\to\infty}\!\! \Bigl\{\!-g
t_g^{0i}-\f{1}{16\pi}\phi_{\alpha\beta}\pa^\alpha
\pa^\beta\!\phi^{0i} +\f{1}{32\pi}
\phi\gamma_{\alpha\beta}\pa^\alpha
\pa^\beta\!\phi^{0i}\!\Bigr\}d\sigma_i\,. \label{eq16} \ee Keeping
in r.h.s. of Eq.~(\ref{eq16}) only terms quadratic in the field we
find from Eq.~(\ref{eq5})  a contribution of the first term to the
flow density \be -g t_g^{0i}=\f{1}{32\pi}
\Bigl\{\gamma^{0\alpha}\gamma^{i\beta}
\Bigl(\pa_\alpha\phi_\tau^\nu\pa_\beta\phi^\tau_\nu
-\f{\,1\,}{2}\pa_\alpha\phi\pa_\beta\phi\Bigr)\Bigr\}\,.
\label{eq17} \ee A contribution to the flow density of the second
term on the basis of Eq.~(\ref{eq2}) is equal to zero and  the
contribution from the third term is equal to \be
-\f{1}{32\pi}m^2\phi\phi^{0i}=-\f{1}{16\pi}R\phi^{0i}.
\label{eq18} \ee
 Thus, the total density of a gravitational radiation flow   will be determined by value [1]:
\be \f{1}{32\pi}\Bigl[\gamma^{0\alpha}\gamma^{i\beta}
\Bigl(\pa_\alpha\phi_\tau^\nu\pa_\beta\phi^\tau_\nu
-\f{\,1\,}{2}\pa_\alpha\phi\pa_\beta\phi\Bigr)
-m^2\phi\phi^{0i}\Bigr]\,, \label{eq19} \ee which as it is shown
in [1] will lead us to the   positively definite flow of
gravitational energy defined as \be
\f{dJ}{d\Omega}=\frac{\,2\,}{\pi}\!\!
\int\limits^{\infty}_{\omega_{\min}}\!\!d\omega\,\omega^2q
\Bigl\{|T^1_2|^2+\frac{1}{4}|T^1_1-T^2_2|^2+ \frac{m^2}{\omega^2}
(|T^1_3|^2+|T^2_3|^2)+\frac{3m^4}{4\omega^4} |T^3_3|^2\Bigr\}\,,
\label{eq20} \ee here $q=[1-(m^2/\omega^2)]^{1/2}$.

\end{document}